\documentclass[usenatbib]{mnras}
\usepackage{epsfig}
\usepackage{times}
\usepackage{amsmath,amssymb}
\usepackage{xcolor}
\usepackage{tikz}
\graphicspath{{figures/}}
\newcommand{\Msolar}{\mbox{\,$\rm M_{\odot}$}}        
\newcommand{\Rsolar}{\mbox{\,$\rm R_{\odot}$}}        

  \newcommand{\Teff}{\mbox{\,\em T$_{\rm eff}$}}         

  \newcommand{\kmsec}{\,\mbox{$\mbox{km}\,\mbox{s}^{-1}$}}    
  \newcommand{\dg}{\mbox{\,$^{\circ}$}}                  
  \def\simge{\mathrel{\raise1.16pt\hbox{$>$}\kern-7.0pt
    \lower3.06pt\hbox{{$\scriptstyle \sim$}}}}           
  \def\simle{\mathrel{\raise1.16pt\hbox{$<$}\kern-7.0pt
    \lower3.06pt\hbox{{$\scriptstyle \sim$}}}}           
\title[Photometry of hydrogen-deficient binaries]{{\it TESS} uncloaks the secondaries in hydrogen-deficient binaries}

\author[C. S. Jeffery]{
  C. Simon Jeffery\thanks{E-mail:simon.jeffery@armagh.ac.uk} 
 \\
Armagh Observatory and Planetarium, College Hill, Armagh BT61 9DG, N. Ireland, UK
 }
 
\date{Accepted \ldots; Received \ldots; in original form \ldots}
\pagerange{\pageref{firstpage}--\pageref{lastpage}} 
\pubyear{2022} 

\begin{document}
\label{firstpage}
\maketitle

\begin{abstract} 
$\upsilon$\,Sgr is the prototype of four known hydrogen-deficient binary (HdB) systems. These are characterised by a hydrogen-deficient A-type primary, variable hydrogen emission lines, and a normally unseen secondary presumed to be an upper main-sequence star. Orbital periods range from tens of days to 360\,d. {\it TESS} observations of all four HdBs show a flux variation with well-defined period in the range 0.5 -- 0.9\,d, too short to be associated with the supergiant primary, and more likely to be the rotation period of the secondary and associated with a chemical surface asymmetry or a low-order non-radial oscillation. The observed rotation period supports a recent analysis of the $\upsilon$\,Sgr secondary. The observations give a direct glimpse of the secondary in all four systems, and should help to explain how the primary has been stripped to become a low-mass hydrogen remnant.
\end{abstract}


%

\section{Introduction}

The small class of hydrogen-deficient binary stars (HdBs) is characterized by their prototype $\upsilon$ Sgr = HD\,181615/6. 
Known for over a century as a naked-eye single-lined spectroscopic binary with an orbital period of 138 days \citep{campbell99,wilson15},  $\upsilon$ Sgr has an atmosphere dominated by CNO-processed helium \citep{hack63,schoenberner83,kipper12}.
The commonly-held view is that $\upsilon$ Sgr and three similar HdBs formed from intermediate mass binary systems. 
Following a first stage of mass transfer on the red-giant branch, they are now in a second stage of mass transfer which has completely stripped the hydrogen envelope to reveal the CNO-processed helium core \citep{schoenberner83,iben85,iben86b,eggleton06}\footnote{Figs.\,1 and 2 in \citet{iben85} nicely illustrate the evolution of a close binary to yield a low-mass hydrogen-stripped supergiant. Fig.\,2 should have replaced the single extreme helium and R\,CrB stars with HdBs.}.
As such they offer important tests for models of the late evolution of close binary stars. 

Despite or perhaps because of the brightness of the primary star, the mass of the primary and the nature of the  companion to  $\upsilon$ Sgr  has proved elusive. 
A weak signal in the cross-correlation function of phase-resolved ultraviolet spectra of $\upsilon$ Sgr was identified as coming from the secondary star and pointed to a primary mass in excess of $1 \Msolar$ \citep{dudley90}. 
A modern reanalysis of the same data has discounted this result and, with the inclusion of parallax and other improvements, points to a primary mass $0.3^{+0.5}_{-0.2}\Msolar$ \citep{gilkis22}. 
 
Variability in the light from $\upsilon$ Sgr had been suspected since eclipses were reported by \citet{gaposchkin45}, \citet{eggen50} and, in the ultraviolet, by \citet{duvignau79}. 
These variations have been discounted as eclipses following improvements in instrumentation and phase coverage \citep{rao85}.
Instead, a $\sim 20$\,d light variation \citep{malcolm86} has been interpreted as a radial pulsation by \citet{morrison88} and \citet{saio95b}, the latter suggesting a primary mass $\sim 3 \Msolar$.  

Three HdBs similar to $\upsilon$ Sgr have been identified to date. 
The first, HD\,30353 = KS\,Per, was detected with a period of some 360 days \citep{bidelman50,heard55}. 
HDE\,320156 = LSS\,4300 = V1037\,Sco and CPD-58$\dg$2721 = LSS\,1922 = V426\,Car were recognised to be hydrogen-deficient during a survey of luminous OB stars \citep{drilling80,schoenberner84} and to be binaries shortly after \citep{drilling85,jeffery87}. 
Efforts to measure the orbital periods of both systems have not been entirely successful \citep{jeffery87b,frame95}. 
On the other hand, the atmospheres are hydrogen-deficient and N-rich \citep{wallerstein67,schoenberner84,morrison87}, indicative of CNO-processed helium. All show low surface gravities and variable hydrogen emission lines \footnote{\citet{fleming92b} noted  emission-lines in the spectrum of LSS\,1922 over 130 years ago.}. 

KS\,Per, LSS\,4300 and LSS\,1922 also show photometric variability in the manner of $\upsilon$\,Sgr, with characteristic periods $\sim$ 30, 20 and 15--20\,d respectively \citep{osawa63,morrison87b,morrison87c,jones89,frame95}. 
Difficulties are that the variability appears not to be regular and covers periods that are challenging for long-term ground-based photometry.  
This is a feature common to other luminous hydrogen-deficient stars, including extreme helium stars and R\,CrB variables \citep{jeffery08.ibvs}, and is likely attributable to the extreme non-adiabacity and non-linearity of low-order strange-mode pulsations \citep{saio95b,montanes02.thesis}. 

Continuous long-term photometric monitoring of variable stars from 
space with {\it e.g.} the {\it COROT}, {\it MOST}, {\it Kepler} and {\it TESS}
spacecraft has transformed our understanding of stellar variability. 
It was of considerable interest when the  {\it TESS} lightcurve for KS\,Per became available (Sector 19) and showed an unexpected variation with a period of some 0.8\,d. 
This was too short to be physically associated with the supergiant primary, and therefore had to be due to the unseen secondary. 
As the mission proceeded, similar discoveries were made for  LSS\,4300 and LSS\,1922, offering a new tool to explore the dimensions of these intriguing binaries. 

With the release of {\it TESS} data for $\upsilon$\,Sgr (Sector 54) came the  realisation that all four HdBs show regular short-period variations of 0.5 -- 0.9\,d. 
Why should the companions show such behaviour and why should they be so similar?
This letter describes the observations (\S\,2) and analyses (\S\,3), and discusses possible causes (\S\,4).

\begin{figure}
    \centering
\includegraphics[width=88mm,angle=0,clip=true]{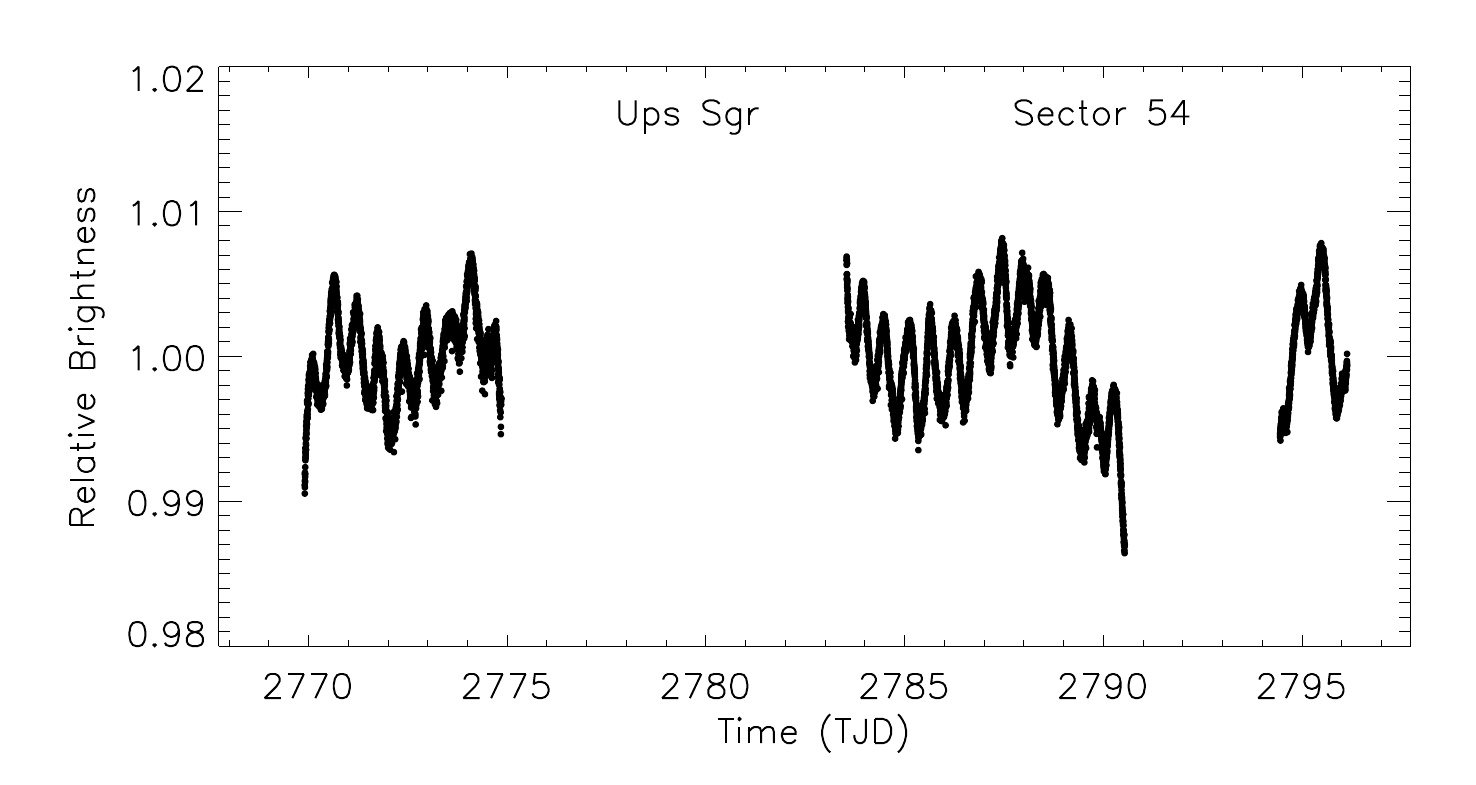}
\includegraphics[width=88mm,angle=0,clip=true]{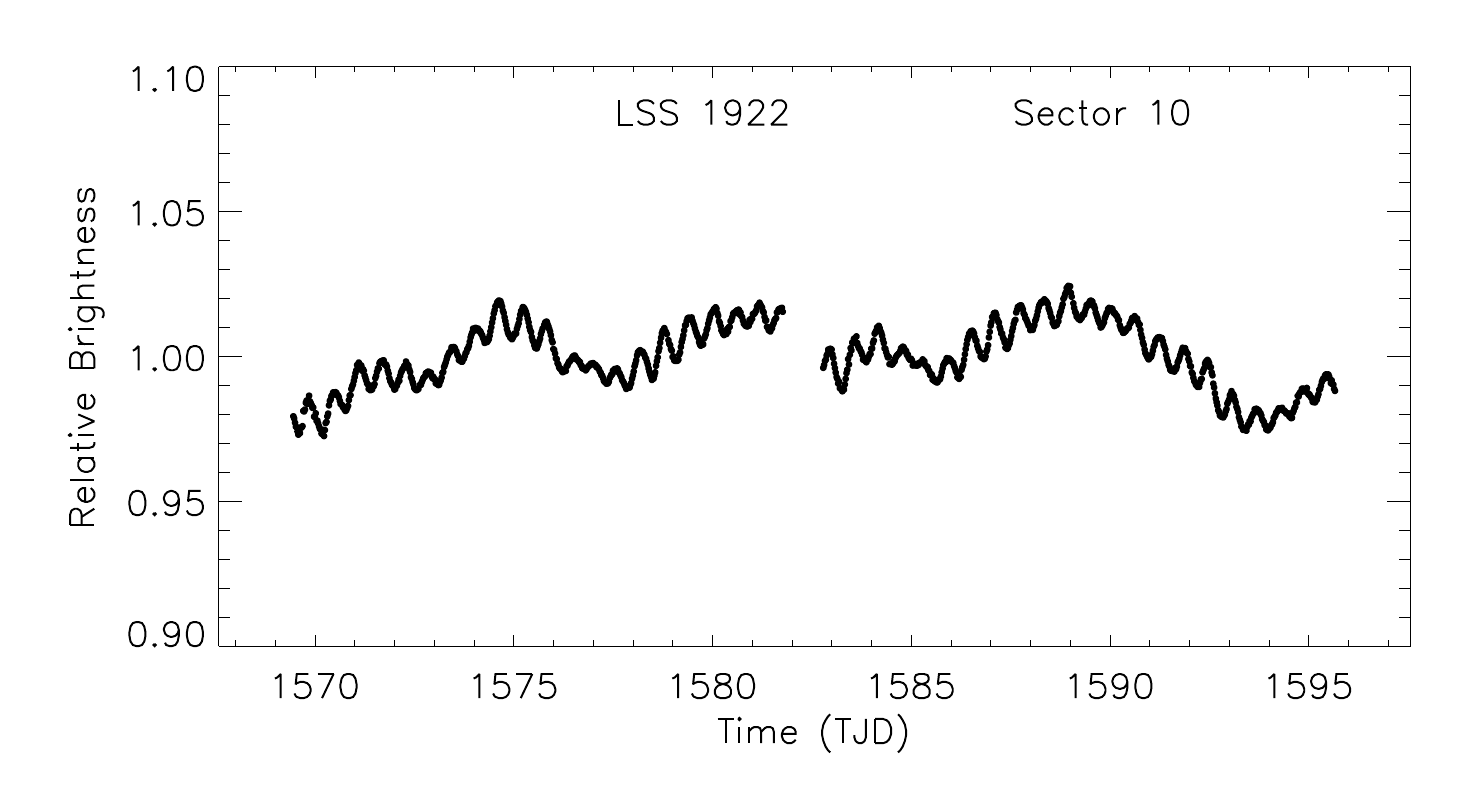}
\includegraphics[width=88mm,angle=0,clip=true]{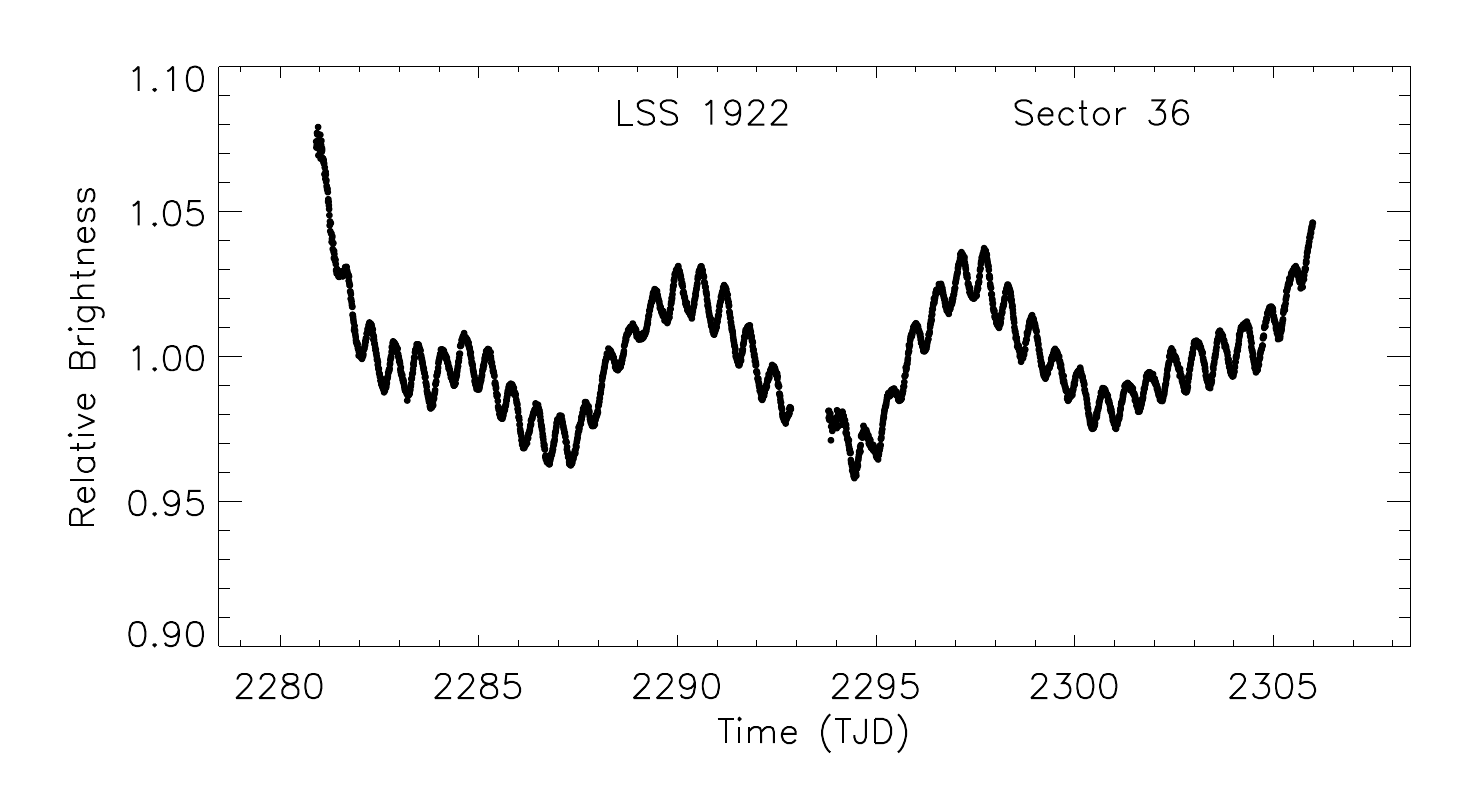}
\includegraphics[width=88mm,angle=0,clip=true]{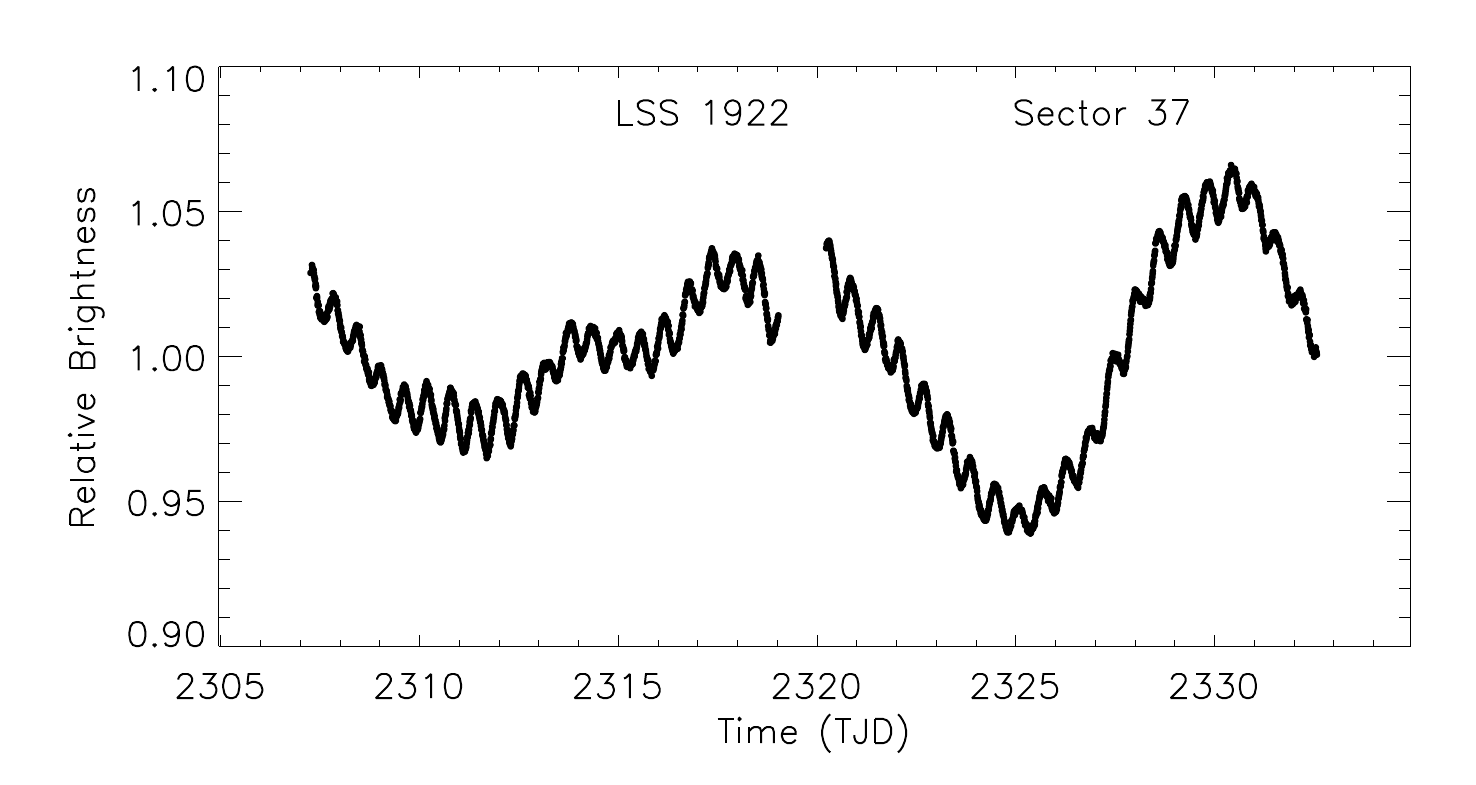}
    \caption{TESS lightcurves by sector for the hydrogen-deficient binaries $\upsilon$\,Sgr and LSS\,1922 (= CPD$-58^{\circ}2721$). Flux is shown relative to the mean brightness over the sector. Time is shown in days since the start of the {\it TESS} mission, ${\rm TJD} = {\rm JD} - 2\,457\,000.0$}
    \label{f:lc1}
\end{figure}
\begin{figure}
    \centering
\includegraphics[width=88mm,angle=0,clip=true]{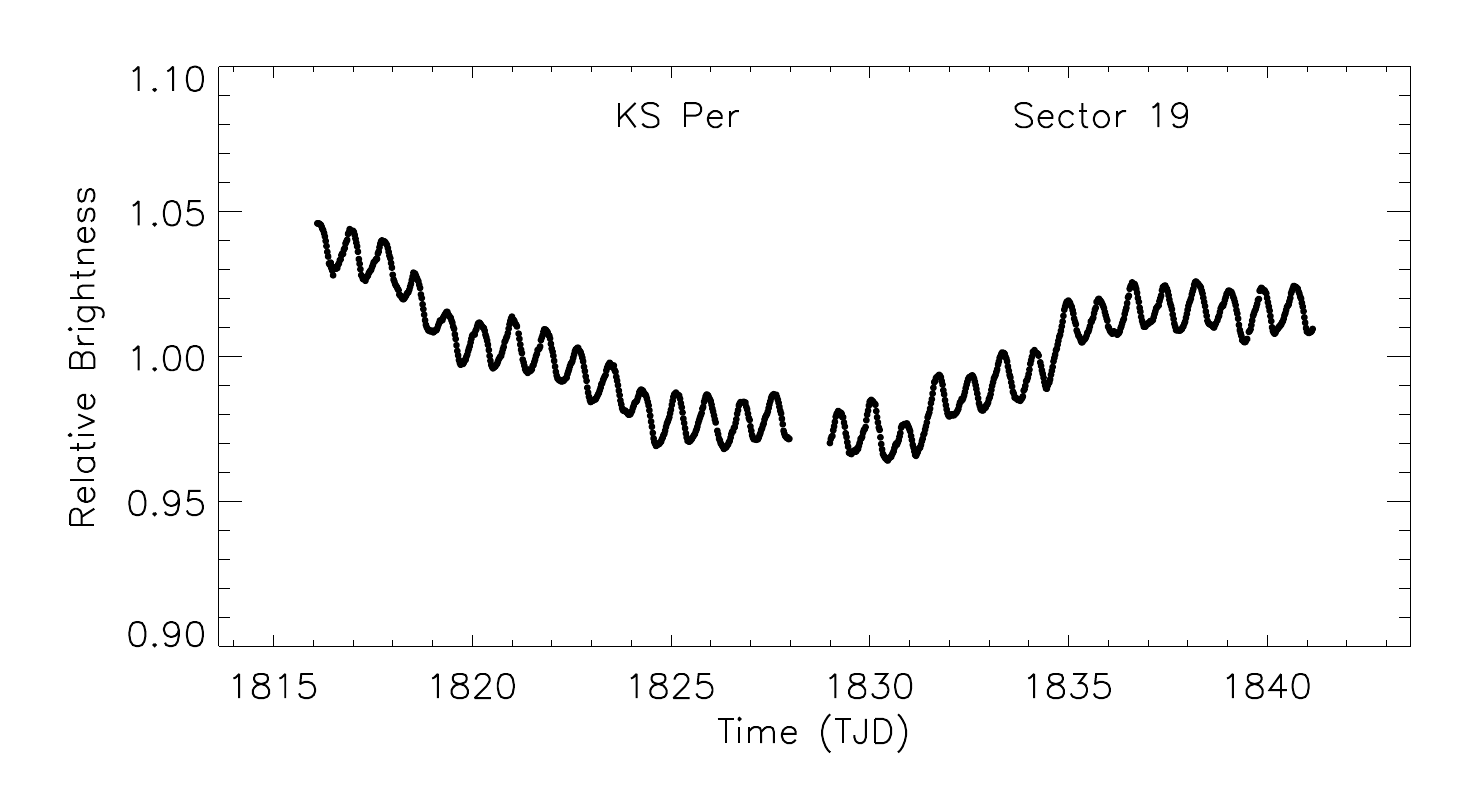}
\includegraphics[width=88mm,angle=0,clip=true]{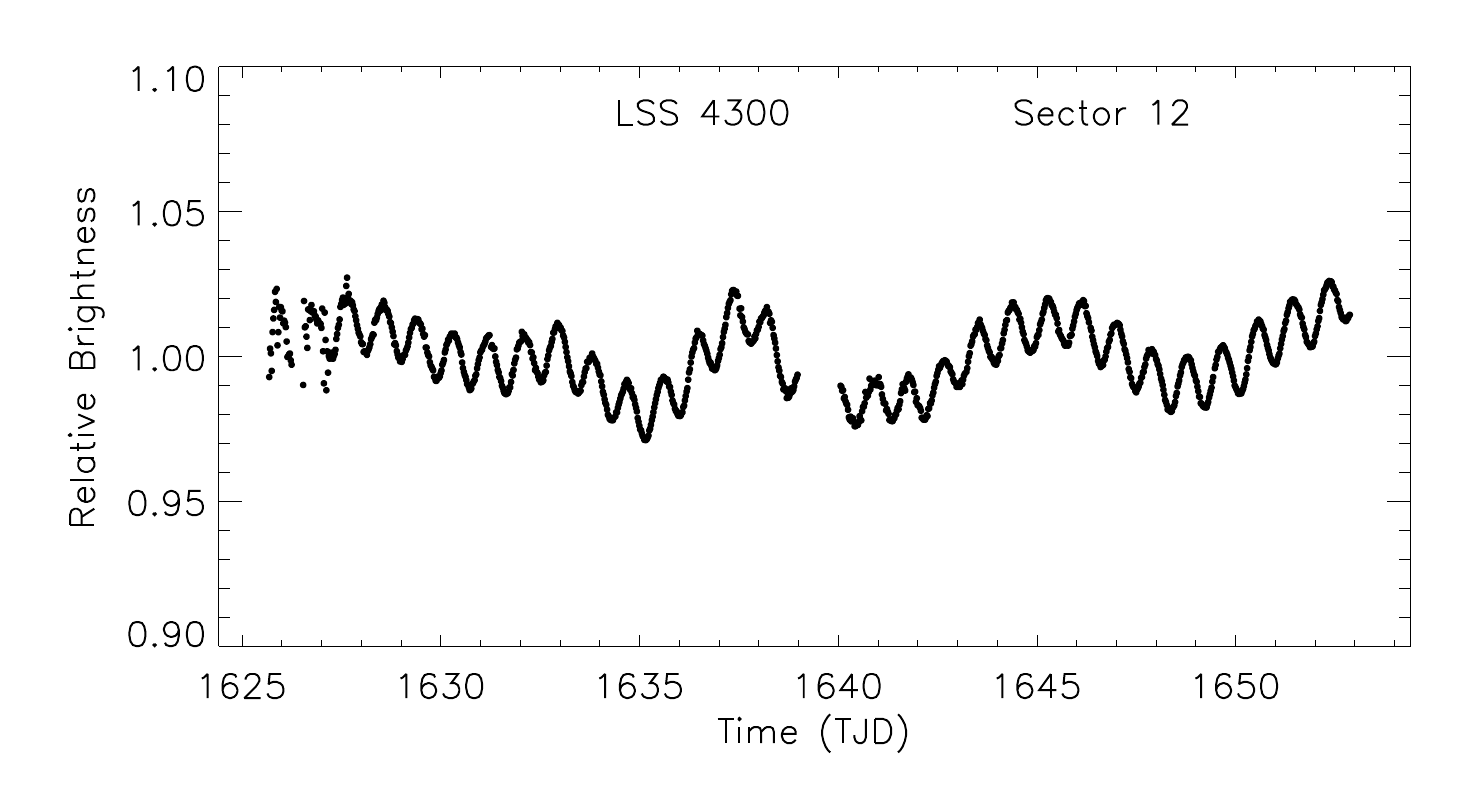}
\includegraphics[width=88mm,angle=0,clip=true]{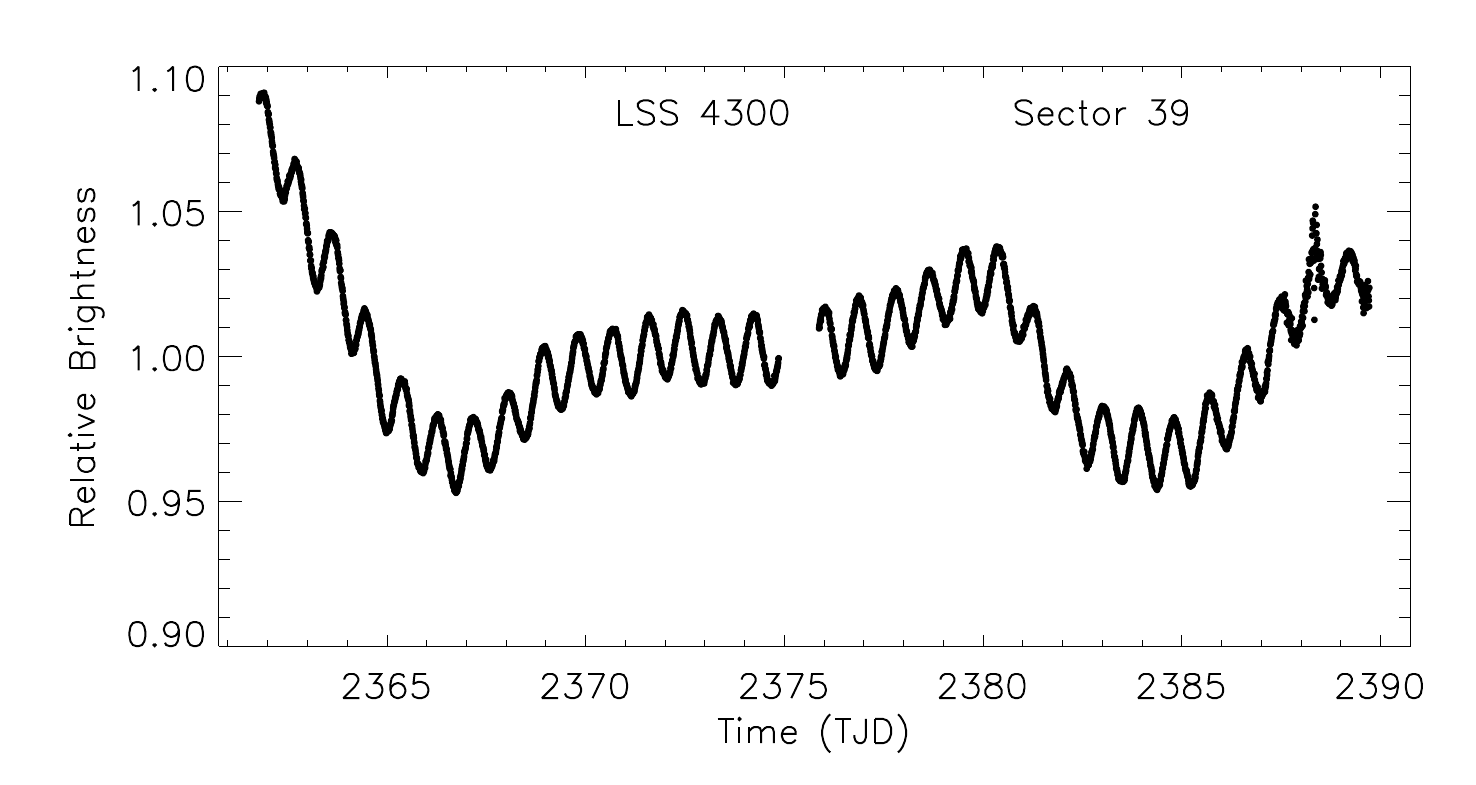}
    \caption{As Fig.~\ref{f:lc1} for KS\,Per and LSS\,4300 (= HDE\,320156).}
    \label{f:lc2}
\end{figure}

\begin{figure}
    \centering
\includegraphics[width=88mm,angle=0,clip=true]{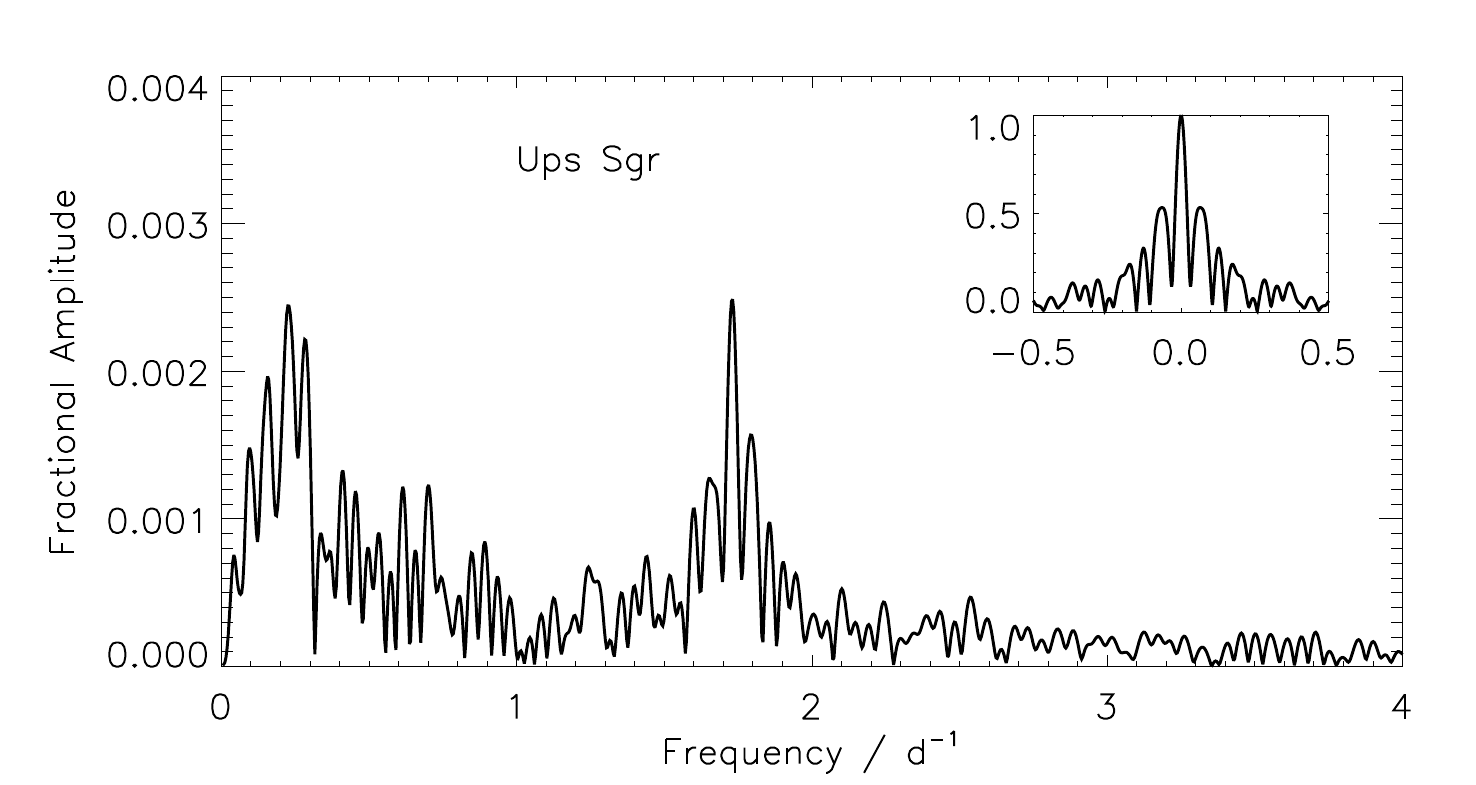}
\includegraphics[width=88mm,angle=0,clip=true]{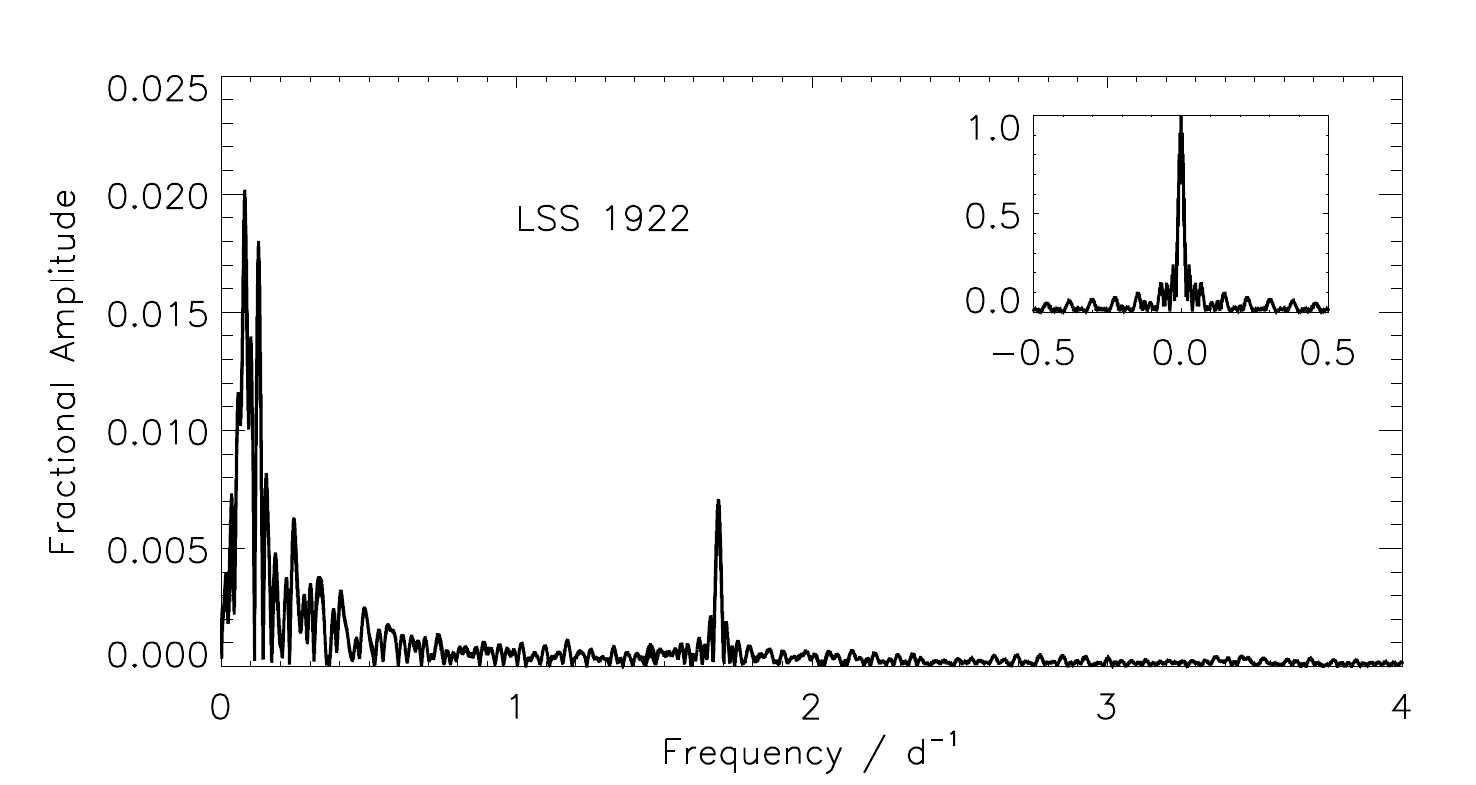}
\includegraphics[width=88mm,angle=0,clip=true]{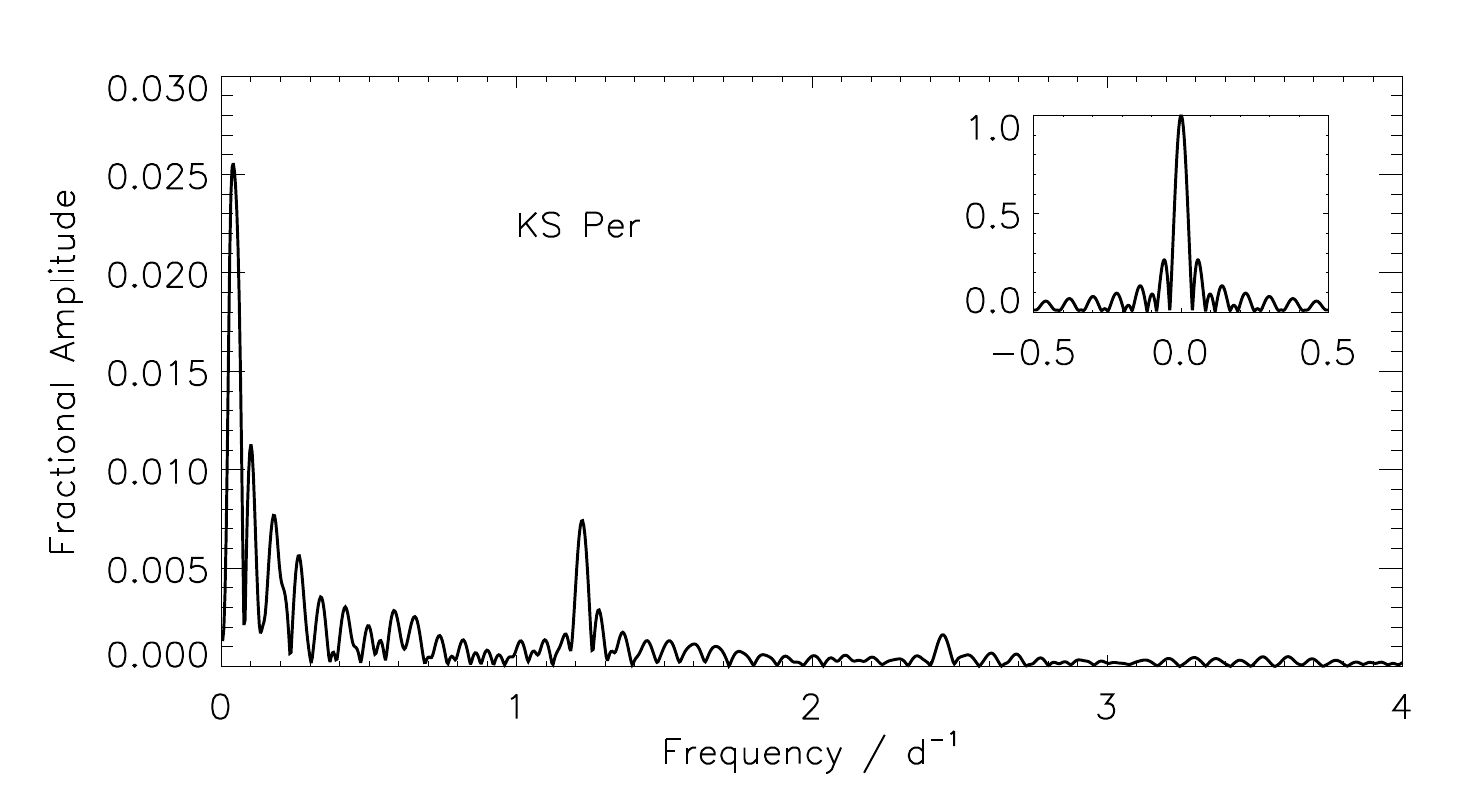}
\includegraphics[width=88mm,angle=0,clip=true]{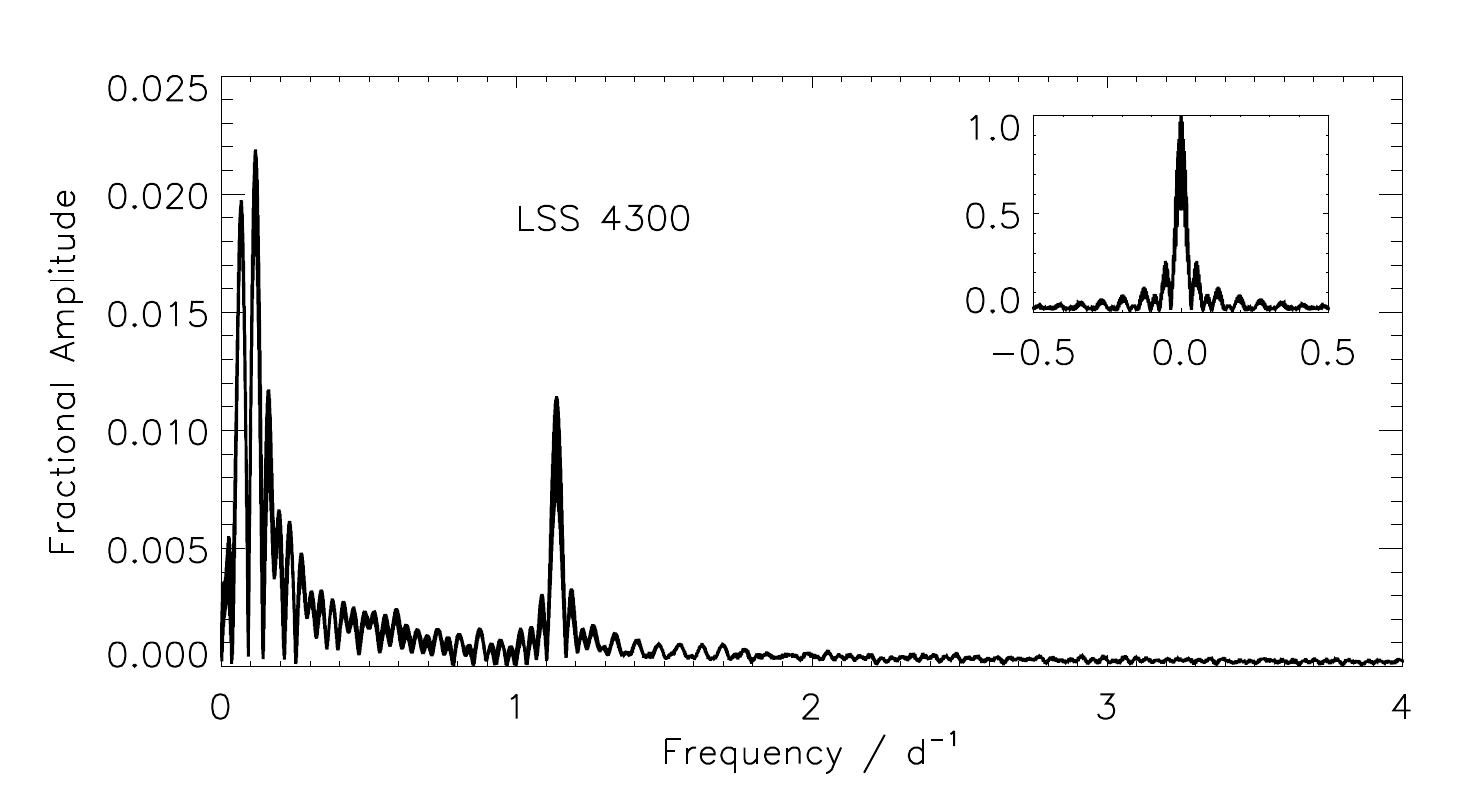}
    \caption{Frequency amplitude spectra derived from the cumulative TESS lightcurve for each of the hydrogen-deficient binaries shown in Figs.~\ref{f:lc1} and \ref{f:lc2}. The window function for each light curve is inset to each panel. }
    \label{f:ps}
\end{figure}

\begin{table}
    \centering
    \begin{tabular}{llcr}
    \hline
    Star & TIC & Sector & Cadence (s) \\
    \hline
$\upsilon$ Sgr & 334020482 & 54 &  120 \\[2mm]
KS\,Per & 367540935 &19 & 1800 \\[2mm] 
 LSS\,4300 & FFI & 12  & 1800 \\
    &           & 39 & 600 \\[2mm]
 LSS\,1922 & FFI & 10  &  1800\\
    &           &  36  & 600 \\
    &           &  37  & 600 \\
\hline
    \end{tabular}
    \caption{Input catalogue numbers (TIC), sectors and cadences for HdBs observed with {\it TESS}. `FFI' indicates that the light curve was recovered from  full-frame images.} 
    \label{t:tess}
\end{table}

\begin{table}
    \centering
    \begin{tabular}{llrrr}
    \hline
    Star	&	& $f / {\rm d^{-1}}$	& $P / {\rm d}$	& $a$ \\
    \hline
    $\upsilon$ Sgr	& $f_1$	&1.730	&0.578	& 0.0025 \\ 
                    &    &0.227 & 4.41 & 0.0024 \\[2mm]
				
    KS Per	& $f_1$	&1.221	&0.819	&0.0074\\
	&$2f_1$	&2.443	&0.409	&0.0016\\
	&	&0.039	&25.4	&0.0259\\[2mm]
				
    LSS 1922	&$f_1$	&1.683	&0.594	&0.0071\\
	&	&0.126	&7.96	&0.0180\\
	&	&0.080	&12.6	&0.0202\\[2mm]
				
    LSS 4300	&$f_1$	&1.134	&0.881	&0.0115\\
	&	&0.116	&8.63	&0.0219\\
	&	&0.067	&15.0	&0.0198\\
	\hline
    \end{tabular}
    \caption{Frequencies $f$, periods $P$ and amplitudes $a$ of significant peaks in the amplitude spectra of hydrogen-deficient binaries. Entries labelled $f_1$, $f_2$ are considered safe and also are likely to arise in the unseen secondary. Unlabelled entries may partially arise from long-period pulsation in the primary, but are strongly contaminated by systematics from the {\it TESS} spacecraft. Amplitudes are given as a fraction of the mean brightness. }
    \label{t:freq}
\end{table}

\section{TESS lightcurves}

{\it TESS} lightcurves for all four HdBs have been downloaded from the Barbara A. Mikulski Archive for Space Telescopes (MAST) using the package {\sc lightkurve} \citep{lightkurve18}. 
{\it TESS} input catalogue (TIC) numbers, sectors and cadences for each dataset are shown in Table\,\ref{t:tess}. 
Data for LSS\,4300 and LSS\,1922 were extracted from Full-Frame Images using {\sc lightkurve}. 

 Data points more than 4$\sigma$ from the mean were removed. A high-pass Gaussian filter with full-width-half-maximum 10\,d removed some low-frequency instrumental characteristics, but care was taken not to remove all low-frequency information.  
 Data were normalised to the mean flux within each sector.
 The resulting lightcurves are shown in Figs.\,\ref{f:lc1} and \ref{f:lc2}.
 
 Variability on timescales of 5 -- 20 d is evident in all four systems, with no indication that it is regular. 
 The signature of a periodic variation on timescales of 0.5 -- 0.9\, d is clear. 
 At first inspection, the amplitude is roughly constant within a sector, with some cycles having lower amplitude. 
 There are sector-to-sector differences in amplitude; these are associated with the spacecraft pointing and how the stellar image is sampled at different locations on the spacecraft detectors. 

\section{Frequency Analysis}

Observations for each star for all sectors observed have been combined without further modification, apart from subtracting the overall mean. 
The Lomb-Scargle frequency-amplitude spectrum for each star has been computed using the method of \citet{press89.period} and is shown in Fig.\,\ref{f:ps}. 
The window function for each light curve is also shown, and reflects the frequency resolution associated with each star. 
The frequencies, periods and amplitudes of the most significant peaks are shown in Table\,\ref{t:freq}. 
A peak is considered significant if it exceeds 4 times the mean local background in the power spectrum. 
In every case a significant peak $f_1$ is found between 1 and 2 cycles per day. 
In the case of KS\,Per, the first harmonic $2f_1$ is also detected, indicating that the variation is slightly non-sinusoidal.  
There are no other significant peaks in any of the frequency-amplitude spectra between 4 and $24\,{\rm d}^{-1}$. 
The low frequency domain looks like red noise, although that noise includes the irregular 20 -- 30\,d pulsations.
The highest peaks in the red-noise domain ($f < 0.3\,{\rm d}^{-1}$) are reported in Table\,\ref{t:freq}; they are unlikely to be real and are not labelled.  

\section{Discussion}

{\it What is the cause of the 0.5 -- 0.9\,d periodic variability in the light curve of the HdBs? }
 
Oscillations in the primary star may be ruled out. The primaries have low surface gravities and hence large radii. 
For effective temperatures $\sim 10\,000 - 14\,000$\,K \citep{dudley93} and surface gravities $\log g / {\rm cm\,s^{-2}} \sim 1 - 2$, the fundamental radial mode would have a period $\simge 3.10^6$\,s or $\simge 30$\,d  \citep[Figs. 2 and 3]{jeffery16a}. 
It would have to a very high-order p-mode to have a period $<1$\, d, and would be unprecedented for such a supergiant.

Oscillations in the secondary star are a possibility. 
If an upper main-sequence star, candidate processes include low-order radial and non-radial p-mode oscillations, as in $\beta$ Cepheids, or non-radial g-mode oscillations, as in slowly-pulsating B (SPB) stars. 
Neither is satisfactory.  
$\beta$ Cepheid pulsations give characteristic multi-periodic amplitude-frequency spectra with frequencies in the range 3 - 16\,d$^{-1}$ \citep{stankov05,labadie20}, too high to reconcile with the HdBs. 
Most SPBs have frequencies $<1\,{\rm d}^{-1}$, although exceptions exist, and are invariably multi-periodic.  

Balona (2022: TASOC preprint) has provided an atlas of light curves and periodograms for variable stars observed by {\it TESS}. 
Additional examples from {\it Kepler} are provided by \citet{balona15,balona16}. 
Amongst these, the only qualitatively close match to the periodograms shown in Fig.\,\ref{f:ps} is classified ROT: ``A and B stars showing an isolated low-frequency, low amplitude peak at less than $<4\,{\rm d}^{-1}$, consistent with what one might expect for the rotational frequency. The harmonic of the peak is often present'' 
(Balona 2022). 

{\it We conjecture that short period variability observed in  HdBs corresponds to  the rotation period of the secondary.}
Rotational modulation must be associated with a simple asymmetry of the secondary surface. 

One possibility is a chemical inhomogeneity  associated with a magnetic field\footnote{The $\upsilon$\,Sgr primary shows a weak magnetic field of some tens of Gauss \citep{hubrig22}; far-ultraviolet spectropolarimetry could in principle detect a magnetic field in the secondary.}. 
{\it Kepler} light curves of several magnetic chemically peculiar (mCp) stars are shown by \citet{hummerich18}, with several being sinusoidal or nearly sinusoidal. 
The mCp sample shows spectral types in the range B7 -- A5, and rotation frequencies in the range 0.1 -- 1.1 d$^{-1}$. 
These are low compared with the HdBs, but one could imagine such a star in a binary accreting matter and gaining angular momentum during a mass transfer episode. 

A second possibility is a low-order non-radial oscillation. For example, an oscillatory convection mode in the core can resonantly excite a $g$ mode in the envelope if the core is rotating slightly faster than the envelope \citep{lee20,lee21}, producing a sinusoidal light variation with a frequency close to $f_{\rm rot}$ ($l = 1, m=-1$). The envelope could rotate more slowly than the core if, for example, it expands in response to the accretion of material from a secondary, but not if it is also spun up by the addition of angular momentum. 

The spectroscopic detection and analysis of the secondary star in $\upsilon$ Sgr indicates a spectral type B2\,V, a projected surface rotation speed $ v_{2,{\rm rot}} \sin i = 250\pm20\kmsec$ and radius $R_2=2.2\pm0.3\Rsolar$ \citep{gilkis22}. 
Hence, a minimum rotation frequency $f_{2,\rm rot} > 2.25\pm0.35\,{\rm d}^{-1}$ is indicated, in weak agreement  with the {\it TESS} photometric frequency $f_{1} = 1.730\,{\rm d}^{-1}$. 
A $1\sigma$ increase in radius combined with a 1$\sigma$ decrease in projected rotation velocity would substantially reduce this discrepancy ($f_{2,\rm rot} > 1.82\pm0.27\,{\rm d}^{-1}$). 
The first is easily accommodated by the evolution model proposed by \citet{gilkis22}.
A second discrepancy concerns the inferred effective temperature of the secondary \citep[$\Teff = 23\,000\pm2\,000$\,K:][]{gilkis22}, which is high compared with the mCp stars described by \citep{hummerich18}. 
On the other hand rotational modulation with frequencies in the observed range is observed in main-sequence stars with spectral types as early as B1.5 \citep[HD\,253214: ][]{balona16}. 

In principle, the visibility of a regular pulse associated with the secondary star  could provide a means to measure the orbital motion, but success seems unlikely. In the case of $\upsilon$ Sgr, the orbital solution by \citet{gilkis22} gives a difference in pulse time of $\sim 26\pm5$\,s between apogee and perigee, or 0.0005 cycles. This would be a challenge, even using data from 6 consecutive {\it TESS} sectors to cover the full orbital cycle. 

The presence of rotational modulation in light from the secondaries of all four HdBs and with periods in a relatively narrow range is both unexpected and remarkable. 
It points to a strong similarity between all four binary systems and, in particular, their initial states, even though their present-day orbital periods differ widely.
The accepted model for these systems requires two stages of mass transfer from the primary onto the secondary. 
It is likely that at least one of those stages has determined the rotation period of the secondary, probably through tidal interaction during the first stage of mass transfer. 
If the surface inhomogeneity responsible for the flux variation has a magnetic origin, the surface field must have survived or been created during this accretion phase. 

\section{Conclusion}
{\it TESS} photometry of all four known hydrogen-deficient binaries has revealed the presence of a nearly sinusoidal periodic variation  in the range 0.5 -- 0.9\,d with amplitude 0.2 -- 2\%, in addition to the longer-period variability already known. 
The short-period variation is too short to be associated with the hydrogen-deficient A-type supergiant primary, and hence must be associated with the usually unseen companion. 
The inference that the latter have rotation periods in the range 0.5 -- 0.9\,d is supported by observations of the secondary in the $\upsilon$\,Sgr system \citep{gilkis22}.
To exhibit rotation modulation, the surfaces must be illuminated asymmetrically. Possibilities include a chemical surface inhomogeneity associated with a magnetic field, or a low-order non-radial oscillation.

\section*{Acknowledgments}
The author thanks Hideyuki Saio for helpful advice in preparing this paper.  

The author is indebted to the UK Science and Technology Facilities Council via UKRI Grant No. ST/V000438/1 for grant support, and to the Northern Ireland Department for Communities which funds the Armagh Observatory and Planetarium.

This paper includes data collected with the {\it TESS} mission, obtained from the MAST data archive at the Space Telescope Science Institute (STScI). Funding for the {\it TESS} mission is provided by the NASA Explorer Program. STScI is operated by the Association of Universities for Research in Astronomy, Inc., under NASA contract NAS 5–26555.

This research made use of {\sc lightkurve}, a Python package for {\it Kepler} and {\it TESS} data analysis \citep{lightkurve18}.

For the purpose of open access, the author has applied a Creative Commons Attribution (CC BY) license to any Author Accepted Manuscript version arising.

\section*{Data Availability}
All data used for this paper are available from MAST.

\bibliographystyle{mnras}
\bibliography{ehe}
\label{lastpage}
\end{document}